# Rank Awareness in Joint Sparse Recovery

Mike E. Davies, *Member IEEE* and Yonina C. Eldar, *Senior Member, IEEE*

## Abstract

In this paper we revisit the sparse multiple measurement vector (MMV) problem, where the aim is to recover a set of jointly sparse multichannel vectors from incomplete measurements. This problem has received increasing interest as an extension of single channel sparse recovery, which lies at the heart of the emerging field of compressed sensing. However, MMV approximation has origins in the field of array signal processing as we discuss in this paper. Inspired by these links, we introduce a new family of MMV algorithms based on the well-know MUSIC method in array processing. We particularly highlight the role of the rank of the unknown signal matrix $\mathbf{X}$ in determining the difficulty of the recovery problem. We begin by deriving necessary and sufficient conditions for the uniqueness of the sparse MMV solution, which indicates that the larger the rank of $\mathbf{X}$ the less sparse $\mathbf{X}$ needs to be to ensure uniqueness. We also show that as the rank of $\mathbf{X}$ increases, the computational effort required to solve the MMV problem through a combinatorial search is reduced.

In the second part of the paper we consider practical suboptimal algorithms for MMV recovery. We examine the rank awareness of popular methods such as Simultaneous Orthogonal Matching Pursuit (SOMP) and mixed norm minimization techniques and show them to be *rank blind* in terms of worst case analysis. We then consider a family of greedy algorithms that are *rank aware*. The simplest such method is a discrete version of the MUSIC algorithm popular in array signal processing. This approach is guaranteed to recover the sparse vectors in the full rank MMV setting under mild conditions. We then extend this idea to develop a rank aware pursuit algorithm that naturally reduces to Order Recursive Matching Pursuit (ORMP) in the single measurement case. This approach also provides guaranteed recovery in the full rank setting. Numerical simulations demonstrate that the rank aware techniques are significantly better than existing methods in dealing with multiple measurements.

## I. INTRODUCTION

Sparse signal representations provide a general signal model that represents or approximates a signal using a linear combination of a small number of elementary waveforms (called atoms) selected from a large collection (the dictionary). Such models make it possible to solve many ill-posed problems such as source separation, denoising

M. E. Davies is with the Institute for Digital Communication, Edinburgh University, Edinburgh EH9 3JL, U.K. (e-mail: mike.davies@ed.ac.uk). MED acknowledges support of his position from the Scottish Funding Council and their support of the Joint Research Institute in Signal and Image Processing with the Heriot-Watt University as a component of the Edinburgh Research Partnership. This work was supported in part by the SMALL project with financial support of the Future and Emerging Technologies (FET) programme within the Seventh Framework Programme for Research of the European Commission, under FET-Open grant number: 225913.

Y. Eldar is with the Technion—Israel Institute of Technology, Haifa 32000, Israel. Phone: +972-4-8293256, fax: +972-4-8295757, E-mail: yonina@ee.technion.ac.il. She is currently on leave at Stanford, USA. Her work was supported in part by the Israel Science Foundation under Grant no. 1081/07 and by the European Commission in the framework of the FP7 Network of Excellence in Wireless COMmunications NEWCOM++ (contract no. 216715).

and most recently compressed sensing [1], [2] by exploiting the additional sparsity constraint. The key point is that when the signal, $\mathbf{x}$, is sufficiently sparse it can still be uniquely determined from an underdetermined set of measurements $\mathbf{y} = \mathbf{\Phi x}$, where $\mathbf{\Phi} \in \mathbb{R}^{m \times n}$ and $m < n$.

The problem of finding the sparsest $\mathbf{x}$ consistent with a given observation vector $\mathbf{y}$ is known to be NP-hard in general [3], [4] and therefore is presumed to not be solvable in polynomial time. Instead, various suboptimal strategies have been proposed and have been demonstrated, both empirically and theoretically, to have good reconstruction performance in a range of settings. Commonly used strategies are typically based on convex relaxation [5], non-convex local optimisation [6] or greedy search strategies [3], [4], [7], [8], [9].

Encouraged by the potential power of sparse representations, researchers have begun to consider a number of extensions to the basic sparse representation model. These include the multiple measurement vector (MMV) problem [10], [11], [12], [13], as well as other union of subspace models [14], [15] such as block sparsity or tree structured sparsity [15], [16], [17] and blind compressed sensing [18]. These ideas have also been recently expanded to include sub-Nyquist sampling of structured analog signals [19], [20], [21], [22], [23]. As with the single measurement vector problem (SMV) several suboptimal methods for finding a sparse matrix have been proposed, that have polynomial complexity [24], [25], [26], [10], [11], [12], [16], [15], [13]. These approaches are generally straightforward extensions of existing SMV solutions and can be roughly divided into greedy methods, and algorithms based on mixed norm optimization. We will discuss these two classes in Section V. One exception to this is the approach in [12] which reduces the MMV problem to a single channel recovery via a random projection that preserves the sparsity pattern.

A variety of different equivalence results between finding the sparsest solution, the so-called $\ell_0$-problem, and the output of the proposed efficient algorithms have also been derived. In [11] an equivalence result was obtained for a mixed $\ell_{p,1}$ program in which the objective is to minimize the sum of the $\ell_p$-norms of the rows of the estimated matrix whose columns are the unknown vectors. The condition is based on mutual coherence, and turns out to be the same as that obtained from a single measurement problem, so that the joint sparsity pattern does not lead to improved recovery capabilities as judged by this condition. Recovery results for the more general problem of block-sparsity were developed in [15] based on the RIP, and in [16] based on mutual coherence. Reducing these results to the MMV setting leads again to conditions that are the same as in the single measurement case. An exception is the work in [27], [28], [13] which considers average case performance assuming that $\mathbf{X}$ is generated at random from an appropriate distribution. Under a mild condition on the sparsity and on the matrix $\mathbf{\Phi}$, the probability of reconstruction failure decays exponentially with the number of channels $l$. However, to date, all worst case recovery results have not shown any advantage to the MMV setting over the SMV case. The reason is that in the worst case, the matrix $\mathbf{X}$ may be comprised of a single repeated vector $\mathbf{x}$, in which case effectively the MMV and SMV problems becomes identical, hence the worst case results are not capable of improving the recovery guarantees.

As noted above, one approach to demonstrate the advantage of the MMV formulation is to use an average case analysis where $\mathbf{X}$ is generated at random. Since repeated columns are not likely to occur, this strategy allows for improved recovery guarantees. In this paper, we concentrate on worst-case performance, as in the bulk of prior work on MMV problems. We show that we can break the worst-case analysis bottleneck by exploiting the rank of the matrix $\mathbf{X}$. Although in the case of a rank-one matrix we cannot do better than in SMV recovery, this is no longer true when $\mathbf{X}$ has higher rank. In particular, when the rank of $\mathbf{X}$ is equal to $k$, we highlight the fact that it can be recovered exactly from the measurements $\mathbf{Y}$ under a mild condition on $\mathbf{\Phi}$ using polynomial time algorithms from only $m = k + 1$ measurements per signal [29] based on the MUSIC algorithm popular in array signal processing [30] . Clearly this is a big advantage over the SMV problem which cannot guarantee perfect recovery for general $\mathbf{\Phi}$ and all $\mathbf{x}$ with such few measurements. Even using combinatorial algorithms, recovery of all $\mathbf{x}$ is possible only if the number of measurements is at least $2k$.

Interestingly, the links between sparse representations and array signal processing were brought forward early on in the sparse reconstruction literature [6] and the MMV problem in particular has origins in the field of array signal processing. However, the algorithmic connections and particularly the role of the rank of $\mathbf{X}$, appear to have been missed.

The main contribution of this paper is to demonstrate how the rank of $\mathbf{X}$ can be exploited in order to improve MMV recovery results in the worst-case setting. We begin by examining the conditions for which the equation $\mathbf{Y} = \mathbf{\Phi X}$ has a single sparse solution and, in Section III, we derive a necessary and sufficient condition for uniqueness and show that it depends directly on the rank of $\mathbf{X}$: the larger the rank the less sparse $\mathbf{X}$ needs to be to still ensure uniqueness. Similarly, in Section IV, we show that the computational effort required to find the unique sparse solution through a combinatorial search is also dependent on the rank of $\mathbf{X}$, and is reduced when the rank increases. In Sections V and VI we turn to discuss polynomial time recovery algorithms. We begin by showing that the common MMV methods are not rank aware, namely, they do not efficiently exploit the rank of $\mathbf{X}$ to improve recovery results. In particular, in the full rank case in which the rank of $\mathbf{X}$ is equal to $k$ (which is the largest it can be) we show that almost all previous greedy and optimization-based MMV algorithms do not provide recovery for the worst-case choice of $\mathbf{X}$ from $k + 1$ measurements. Moreover, independent of the rank of $\mathbf{X}$, one can find examples that perform almost as badly as worst case SMV problems. We proceed to propose some rank aware algorithms inspired by the popular MUSIC technique, whose behavior improves with increasing rank of $\mathbf{X}$, and are proven to provide exact recovery for all choices of $\mathbf{X}$ when the rank is equal to $k$, and the number of measurements is $k + 1$ (under mild conditions on $\mathbf{\Phi}$). Finally, in Section VII we present several simulations demonstrating the behavior of these different methods.

## II. Notation and Problem Formulation

### A. Notation

Throughout this paper a coefficient vector $\mathbf{x}$ is said to be $k$-sparse if the size of the support of $\mathbf{x}$ is no larger than $k$: $|\operatorname{supp}(\mathbf{x})| \leq k$. We define the support of a collection of vectors $\mathbf{X} = [\mathbf{x}_1, \ldots, \mathbf{x}_l]$ as the union over all the individual supports:

$$\operatorname{supp}(\mathbf{X}) := \bigcup_i \operatorname{supp}(\mathbf{x}_i). \tag{1}$$

 

A matrix $\mathbf{X}$ is called $k$ joint sparse if $|\operatorname{supp}(\mathbf{X})| \leq k$. In other words, there are at most $k$ rows in $\mathbf{X}$ that contain nonzero elements. The set of $k$ joint sparse matrices of size $n \times l$ is defined as

$$\mathcal{J}_k := \{\mathbf{X} \in \mathbb{R}^{n \times l} : |\operatorname{supp}(\mathbf{X})| \leq k\}. \tag{2}$$

We make use of the subscript notation $\mathbf{x}_\Omega$ to denote a vector that is equal to some $\mathbf{x}$ on the index set $\Omega$ and zero everywhere else. Denoting by $|\Omega|$ the cardinality of $\Omega$, the vector $\mathbf{x}_\Omega$ is $|\Omega|$-sparse. For index sets that are sequences of indices, e.g. $1, \ldots, k$ we use the Matlab style notation $1 : k$.

We say that the support of $\mathbf{x}$ lies within $\Omega$ whenever $\mathbf{x}_\Omega = \mathbf{x}$. For matrices the subscript notation $\mathbf{\Phi}_\Omega$ will denote a submatrix composed of the columns of $\mathbf{\Phi}$ that are indexed in the set $\Omega$, while the notation $\mathbf{X}_{\Omega,:}$ denotes a row-wise submatrix composed of the rows of $\mathbf{X}$ indexed by $\Omega$. We denote the $i$th column of a matrix, $\mathbf{\Phi}$, by $\boldsymbol{\phi}_i$, and use $\mathcal{N}(\mathbf{\Phi})$ for the null space and $\mathcal{R}(\mathbf{\Phi})$ for the range of the matrix $\mathbf{\Phi}$.

Throughout the paper, we also require a variety of different norms. We denote by $\|\mathbf{x}\|_p$, $p \geq 1$ the usual $\ell^p$ norms, and by $\|\mathbf{x}\|_0$ the $\ell^0$ quasi-norm that counts the number of non-zero elements of $\mathbf{x}$ so that $\|\mathbf{x}\|_0 = |\operatorname{supp}(\mathbf{x})|$. For matrices we define the $\ell_{p,q}$ norms as:

$$\|\mathbf{X}\|_{p,q} := \left(\sum_i \|\mathbf{X}_{i,:}\|_p^q\right)^{1/q} \tag{3}$$

where, with slight abuse of notation, we also consider the quasi-norms with $p = 0$ such that $\|\mathbf{X}\|_{0,q} = |\operatorname{supp}(\mathbf{X})|$ for any $q$.

### B. MMV sparse recovery problem

We are interested in solving the sparse recovery problem associated with Multiple Measurement Vectors (MMV), in which the goal is to recover a jointly sparse matrix $\mathbf{X}$ of size $n \times l$ from $m < n$ measurements per channel. Here $l$ denotes the number of channels, or signals. This is a generalization of the standard Single Measurement Vector (SMV) problem that has been examined in detail, e.g. [31], [7]. Formally, the problem is defined as follows:

**Definition 1** (MMV sparse recovery problem). *Given* $\mathbf{Y} \in \mathbb{R}^{m \times l}$ *and* $\mathbf{\Phi} \in \mathbb{R}^{m \times n}$ *with* $m < n$ *find:*

$$\hat{\mathbf{X}} = \operatorname*{argmin}_{\mathbf{X}} |\operatorname{supp}(\mathbf{X})| \ \textit{s.t.} \ \mathbf{\Phi}\mathbf{X} = \mathbf{Y}. \tag{4}$$

The SMV problem can be recovered as a special case with $l = 1$.

Throughout the paper we will assume that the dictionary $\mathbf{\Phi}$ has unit norm columns, $\|\boldsymbol{\phi}_i\|_2 = 1$. we will also focus on the ideal model in which the measurements $\mathbf{Y}$ are noiseless, and $\mathbf{X}$ is strictly sparse. We do not treat the case in which the signals depart from the exact sparse model or any possible measurement errors, although such issues are clearly very important in practice.

### III. MMV Uniqueness

While the MMV problem reduces to the SMV one when each observation vector $\mathbf{Y}_{i,:}$ is colinear, in general, the additional measurement vectors should provide further information to the problem and make it easier to determine the support set. The question is: by how much? In this section we will focus on uniqueness conditions and see how the rank of $\mathbf{X}$, or alternatively that of $\mathbf{Y}$, can be used to improve the uniqueness conditions. In later sections we will use the rank to develop more efficient recovery algorithms with improved worst-case guarantees over the SMV setting.

Sufficient conditions for the uniqueness of the MMV sparse recovery problem were given in [11]. Here we complete this result by showing that such a condition is also necessary.

For the SMV problem it is well known that a necessary and sufficient condition for the measurements $\mathbf{y} = \mathbf{\Phi}\mathbf{x}$ to uniquely determine each $k$-sparse vector $\mathbf{x}$ is given by

$$k < \frac{\text{spark}(\mathbf{\Phi})}{2} \tag{5}$$

where the *spark* of $\mathbf{\Phi}$ is defined as the smallest number of columns of $\mathbf{\Phi}$ that are linearly dependent. Since $\text{spark}(\mathbf{\Phi}) \leq m + 1$, we have immediately that $m \geq 2k$, namely, at least $2k$ measurements are needed to ensure uniqueness in the SMV case for all possible choices of $\mathbf{x}$. This also defines a necessary and sufficient condition for the general MMV sparse recovery problem, since one instance of this problem is any SMV problem replicated $l$ times [11].

Chen and Huo [11] showed that when $\text{rank}(\mathbf{Y}) > 1$ the sufficient condition for uniqueness in the MMV sparse recovery problem can be relaxed by exploiting the rank of $\mathbf{Y}$, as incorporated in the following theorem.

**Theorem 1** (Chen and Huo [11]). *A sufficient condition from the measurements* $\mathbf{Y} = \mathbf{\Phi}\mathbf{X}$, $|\text{supp}(\mathbf{X})| = k$, *to uniquely determine the jointly sparse matrix* $\mathbf{X} \in \mathcal{J}_k$ *is*

$$|\text{supp}(\mathbf{X})| < \frac{\text{spark}(\mathbf{\Phi}) - 1 + \text{rank}(\mathbf{Y})}{2}. \tag{6}$$

This condition was further shown in [12] to hold even in the case where the are infinitely many vectors $\mathbf{y}_i$. A direct consequence of this result is that matrices $\mathbf{X}$ which result in matrices $\mathbf{Y}$ with larger rank, can be recovered from fewer measurements. Alternatively, matrices $\mathbf{X}$ with larger support can be recovered from the same number of measurements. Since $\text{rank}(\mathbf{X}) \leq k$, it is obvious that $\text{rank}(\mathbf{Y}) \leq k$. When $\text{rank}(\mathbf{Y}) = k$ and $\text{spark}(\mathbf{\Phi})$ takes on its largest value of $m + 1$, the condition (6) becomes $m \geq k + 1$. Therefore, in this best-case scenario, only $k + 1$ measurements per signal are needed to ensure uniqueness. This is much lower than the value of $2k$ obtained in the SMV setting.

Chen and Huo note that it would also be interesting to bound $k$ in terms of $\text{rank}(\mathbf{X})$ instead of $\text{rank}(\mathbf{Y})$. Naturally we have that $\text{rank}(\mathbf{Y}) \leq \text{rank}(\mathbf{X})$. As we show in the next lemma, we can actually replace $\text{rank}(\mathbf{Y})$ by $\text{rank}(\mathbf{X})$ in the condition (6):

**Lemma 1.** *The sufficient condition of* (6) *is equivalent to*

$$|\text{supp}(\mathbf{X})| < \frac{\text{spark}(\mathbf{\Phi}) - 1 + \text{rank}(\mathbf{X})}{2}. \tag{7}$$

*Proof:* Since $\text{rank}(\mathbf{Y}) \leq \text{rank}(\mathbf{X})$, (6) automatically implies (7). For the reverse direction, suppose that (7) holds. Since $\mathrm{rank}(\mathbf{X}) \leq |\mathrm{supp}(\mathbf{X})|$, we have from (7) that

$$|\mathrm{supp}(\mathbf{X})| < \mathrm{spark}(\boldsymbol{\Phi}) - 1. \tag{8}$$

So $\boldsymbol{\Phi}_\Omega$ , $\Omega = \mathrm{supp}(\mathbf{X})$, must be full rank. This implies that $\mathrm{rank}(\mathbf{X}) = \mathrm{rank}(\mathbf{Y})$ and consequently (7) implies (6). ∎

We now show that both (7) and (6) are necessary and sufficient for uniqueness in the MMV problem.

**Theorem 2.** *Condition* (7)*, or equivalently* (6)*, is a necessary and sufficient condition for the measurements* $\mathbf{Y} = \boldsymbol{\Phi}\mathbf{X}$ *to uniquely determine the jointly sparse matrix* $\mathbf{X} \in \mathcal{J}_k$.

An immediate consequence of the theorem is that in order to have a unique $\mathbf{X}$ with support set of size $k$ when $\mathbf{X}$ has full rank (i.e. $\mathrm{rank}(\mathbf{X}) = k$), it is enough to take $m = k + 1$ measurements, as long as $\mathrm{spark}(\Phi) \geq k + 2$, namely, that every set of $k + 1$ columns of $\boldsymbol{\Phi}$ are linearly independent.

Interestingly the $m \geq k + 1$ bound also occurs in the SMV case, where for almost all dictionaries $\boldsymbol{\Phi}$ almost all $k$ sparse vectors $\mathbf{x}$ are uniquely determined by $\mathbf{y} = \boldsymbol{\Phi}\mathbf{x}$ if $m \geq k + 1$ (for apropriately defined measures) - see [14, Theorem 2.6]. The key difference in the MMV scenario is that the condition on $\mathbf{X}$ that guarantees recovery using only $k + 1$ measurements is readily testable from the observed data, namely that $\mathrm{rank}(\mathbf{Y}) = k$.

*Proof:* Sufficiency follows immediately from Lemma 1 and Theorem 1.

To show necessity we will show that $2k \geq \mathrm{spark}(\boldsymbol{\Phi}) - 1 + \tau$ implies that there exists an $\mathbf{X} \in \mathbb{R}^{n \times l}$ with $\mathrm{rank}(\mathbf{X}) = \tau$ that is not uniquely determined by $\mathbf{Y} = \boldsymbol{\Phi}\mathbf{X}$.

Suppose that $2k \geq \mathrm{spark}(\boldsymbol{\Phi}) - 1 + \tau$. Then there exists a support set $T$, $|T| = 2k - \tau + 1$ and a vector $\mathbf{v} \neq 0$ such that $\boldsymbol{\Phi}_T \mathbf{v} = 0$. Let $\mathbf{V} = [\mathbf{v}, \ldots, \mathbf{v}]$ be the matrix consisting of $l$ replications of $\mathbf{v}$. We can now construct an $\mathbf{X}$ with $\mathrm{supp}(\mathbf{X}) \subset T$, as follows:

$$\mathbf{X}_{T,:} = \begin{bmatrix} \begin{array}{|cc|} \hline V_{\{1:k-\tau+1,:\}} \\ \hline \mathbf{I}_{\tau-1} & \mathbf{0} \\ \hline \mathbf{0}_{\{k-\tau+1 \times l\}} \\ \hline \end{array} \end{bmatrix} \tag{9}$$

where $\mathbf{I}_{\tau-1}$ represents the identity matrix of size $\tau - 1$. By construction $\mathbf{X}$ is $k$-joint sparse and $\mathrm{rank}(\mathbf{X}) = \tau$. However we can also define the matrix $\tilde{\mathbf{X}}$, with $\mathrm{supp}(\tilde{\mathbf{X}}) \subset T$, by:

$$\tilde{\mathbf{X}}_{T,:} = \mathbf{X}_{T,:} - \mathbf{V}. \tag{10}$$

By construction, $\tilde{\mathbf{X}}$ is also $k$-joint sparse, and $\boldsymbol{\Phi}\mathbf{X} = \boldsymbol{\Phi}\tilde{\mathbf{X}}$. Therefore, it follows that (7) is also a necessary condition for uniqueness. ∎

Theorem 2 shows that $\mathrm{rank}(\mathbf{X})$ plays an important role in uniqueness of MMV problems, since the rank defines the dimension of the subspace of sparse coefficients that generate observation vectors. In the ensuing sections we will further show that $\mathrm{rank}(\mathbf{X})$ also plays an important role in the performance of joint sparse recovery algorithms.

## IV. MMV Recovery: Exhaustive Search

Our goal now is to show how the rank of $\mathbf{X}$ can be used to improve MMV recovery. We begin by considering the (generally impractical) exhaustive search approach for $\mathbf{X}$ as this most easily allows us to incorporate the rank information. We then discuss the full rank case for which it is straightforward to develop rank-aware algorithms, namely, algorithms that exploit the rank of the measurements, and are guaranteed to recover the true value $\mathbf{X}$ from only $k+1$ measurements. In Section V we show that standard MMV methods are *not* rank aware. Following which, in Section VI, we suggest two rank-aware algorithms that can be applied in the non-full rank scenario as well.

### A. Combinatorial Search

The sparse recovery problem can naturally be cast in terms of a combinatorial optimization. That is we seek to minimize the number of nonzero rows of $\mathbf{X}$ subject to the constraint $\mathbf{\Phi}\mathbf{X} = \mathbf{Y}$. To count the nonzero rows we can consider the $\ell_0$ norm of the vector consisting of arbitrary norms of the rows of $\mathbf{X}$, leading to the problem:

$$\hat{\mathbf{X}} = \underset{\mathbf{X}}{\operatorname{argmin}} \|\mathbf{X}\|_{0,q} \text{ s.t. } \mathbf{\Phi}\mathbf{X} = \mathbf{Y} \tag{11}$$

($q$ arbitrary). If we know $k$ (or a bound on it, say from the uniqueness conditions) then we can alternatively minimize

$$\hat{\mathbf{X}} = \underset{\mathbf{X}}{\operatorname{argmin}} \min_{\Omega, |\Omega| \le k} \|\mathbf{\Phi}\mathbf{X} - \mathbf{Y}\|_{2,2} \tag{12}$$

Other norms are also possible. The choice of the Frobenius norm is popular in array signal processing [32], as it can be related to a maximum likelihood cost function in the presence of Gaussian observation noise. Schemes such as Alternating Projection [33] can be used to search for a *local* minimum of this cost function and have been applied to great effect [29], [34]. Alternatively a *global* minimum for (12) can be found from an exhaustive search through all $\binom{n}{k}$ support sets $\Omega$ with $|\Omega| = k$, until we find the $\mathbf{\Phi}_\Omega$ that minimizes (12).

### B. Full-Rank MMV: MUSIC

Surprisingly, in the case in which $\mathbf{Y}$ has rank equal to $k$, an exact solution can be found using a very simple search algorithm. This method incorporates the rank information of $\mathbf{Y}$ to efficiently determine $\mathbf{X}$. We refer to this scenario as the full rank case, since, we always have that $\operatorname{rank}(\mathbf{Y}) \le \operatorname{rank}(\mathbf{X}) \le k$. The last inequality is a result of the fact that $\mathbf{X}$ is $k$-sparse. The same approach can be extended to reduce the complexity of the general combinatorial search for arbitrary ranks as well.

To obtain a rank-aware algorithm for the case when $\operatorname{rank}(\mathbf{Y}) = k$, we adapt a discrete version of the MUSIC algorithm [30], popular in array signal processing. To the best of our knowledge, the first application of this discrete form of MUSIC to an MMV problem was proposed and analyzed by Feng and Bresler [29], [34] in the context of multiband sampling. This work substantially predates much of the renewed interest in sparse recovery problems. The connection with array processing techniques and rank of the measurements was later exploited in [22] to develop recovery methods for time delay estimation from low rate samples.

 

Since by assumption, $\text{rank}(\mathbf{Y}) = k$, it follows immediately that $\mathcal{R}(\mathbf{Y}) = \mathcal{R}(\boldsymbol{\Phi}_\Omega)$. Therefore, we have full visibility of $\mathcal{R}(\boldsymbol{\Phi}_\Omega)$ in the given measurements which intuitively can be used to ensure recovery of $\mathbf{X}$. The question is how to do this in practice. Here we rely on the MUSIC principle, which exploits the signal subspace properties. As we noted above, every correct column $\boldsymbol{\phi}_i$ of $\boldsymbol{\Phi}$, lies within the range of $\mathbf{Y}$. Assuming that (7) is satisfied, from uniqueness, it follows that these are the only columns contained in $\mathcal{R}(\mathbf{Y})$. Namely, any $\boldsymbol{\phi}_i$ not in the support of $\boldsymbol{\Phi}_\Omega$, will not lie entirely in $\mathcal{R}(\mathbf{Y})$. Consequently, if we project these vectors onto the orthogonal complement of $\mathcal{R}(\mathbf{Y})$, then the projection will contain energy. We can use this observation to develop a concrete method for finding the values of $\boldsymbol{\phi}_i$ that are in $\Omega$.

Specifically, we first calculate an orthonormal basis $\mathbf{U} = \text{orth}(\mathbf{Y})$ for $\mathcal{R}(\mathbf{Y})$ (for example, by using a singular value decomposition (SVD) of $\mathbf{Y}$. Since $\mathbf{U}$ is an orthonormal basis of the range of $\boldsymbol{\Phi}_\Omega$ each of its columns must lie within this subspace. Furthermore, by our assumption that (7) is satisfied, no other column of $\boldsymbol{\Phi}$ can lie within this subspace. Therefore, mathematically, we have that

$$\|(\mathbf{I} - \mathbf{U}\mathbf{U}^T)\boldsymbol{\phi}_i\|_2 = 0, \text{ if and only if } i \in \Omega. \tag{13}$$

If we therefore select the $k$ atoms that minimize $\|(\mathbf{I} - \mathbf{U}\mathbf{U}^T)\boldsymbol{\phi}_i\|_2$, then we are guaranteed to find the correct support set. Since, by assumption, the columns of $\boldsymbol{\Phi}_\Omega$ are linearly independent, we can determine $\mathbf{X}$ as $\mathbf{X} = \boldsymbol{\Phi}_\Omega^\dagger \mathbf{Y}$. As we have already seen, (7) holds as long as $m \geq k+1$, and all sets of $m$ columns of $\boldsymbol{\Phi}$ are linearly independent.

Readers familiar with the array processing literature will recognize the essential ingredients of the MUSIC algorithm. In array processing it is more common to compute the inverse of (13) which is often called the MUSIC pseudo-spectrum.

We have seen that in the full rank case the MUSIC algorithm can be used to fully recover $\mathbf{X}$ from as little as $k + 1$ measurements per signal. The surprising result is that this is true for *any* choice of $\mathbf{X}$, namely, this is a worst case recovery guarantee. Recovery is obtained using a very simple algorithm that is computationally efficient: it requires a single SVD, and computing the norm of several vectors. When $\text{rank}(\mathbf{Y}) < k$, we can use similar ideas to develop a rank-aware algorithm that will improve as the rank of $\mathbf{X}$ increases. We discuss this approach in Section VI-A which leads to a rank aware thresholding method. This technique reduces to the MUSIC algorithm in the full rank scenario.

### C. Reduced-Rank MMV

When $\tau = \text{rank}(\mathbf{Y}) < k$ we can still exploit the geometry through a reduced combinatorial search. If $\tau < k$ then $\mathcal{R}(\mathbf{Y}) \subset \mathcal{R}(\boldsymbol{\Phi}_\Omega)$ and there must exist at least one subset $\gamma \subset \Omega$ such that $|\gamma| = k - \tau$ and $\boldsymbol{\phi}_i \notin \mathcal{R}(\mathbf{Y})$ (Note that typically *all* $k - \tau$ sized support sets $\gamma \subset \Omega$ will satisfy this). Assuming that $\boldsymbol{\Phi}_\Omega$ has full rank, we then have that $\mathcal{R}([\boldsymbol{\Phi}_\gamma, \mathbf{Y}]) = \mathcal{R}(\boldsymbol{\Phi}_\Omega)$. Since $|\gamma| = k - \tau$ we only need to search over subsets of size $k - \tau$. Of course there are 'only' $\binom{n}{k-\tau}$ such support sets, so this search still has exponential complexity unless $k \approx \tau$.

Specifically let $Q(\gamma)$ be an orthonormal basis for $\mathcal{R}([\boldsymbol{\Phi}_\gamma, \mathbf{Y}])$. Then an optimal $\gamma$ can be found by solving:

$$\hat{\gamma} = \underset{\gamma, |\gamma| = k-\tau}{\text{argmin}} \|\boldsymbol{\Phi}_{\gamma^c}(\mathbf{I} - Q(\gamma)Q(\gamma)^T)\|_{0,q} \tag{14}$$

for an arbitrary $q$. The correct support set can then be recovered by considering the full rank problem associated with the augmented measurement matrix $[\mathbf{\Phi}_{\hat{\gamma}}, \mathbf{Y}]$.

Generally the optimal solution to (14) will not be unique and there will typically be multiple solutions associated with the $\binom{k}{k-\tau}$ subsets of the true support. Interestingly, this presence of multiple equivalent minima here suggests that such a problem might be difficult to convexify.

We also note that for special structures of the matrix $\mathbf{\Phi}$, the support of $\mathbf{X}$ can be recovered exactly even when $\mathrm{rank}(\mathbf{Y})$ is smaller than $k$. This is the case, for example, when $\mathbf{\Phi}$ consists of exponentials. In this case, the MUSIC approach can be replaced by the ESPRIT method, which stands for estimation of signal parameters via rotational invariance techniques [35]. However since our emphasis here is on general matrices $\mathbf{\Phi}$, we do not elaborate more on this point.

Motivated by the fact that a higher dimensional range space of $\mathbf{Y}$ in MMV problems plays an important role in the identifiability of $\mathbf{X}$ and can also help reduce the complexity of the combinatorial search we next examine which practical algorithms are able to make use of this rank information and which do not.

## V. Rank-Blind MMV Algorithms

Despite the fact that the rank of $\mathbf{X}$ (and therefore that of $\mathbf{Y}$) plays an important role in the MMV problem, we next show that some of the most popular algorithms are effectively *rank blind*. Namely, they do not allow for perfect recovery in the full rank case, and furthermore the worst case behaviour of such algorithms approaches that of the SMV problem. In practice, as we will see in Section VII, such algorithms often exhibit improved performance when there are multiple measurements, however, they suffer significantly from not properly exploiting the rank information.

### A. Greedy Methods

Several greedy algorithms have been proposed to treat the MMV problem. Two examples are extensions of thresholding and OMP to the multiple measurement case [25], [36], [27], [13], [16]. For $1 \leq q \leq \infty$ they produce a $k$-sparse signal $\hat{\mathbf{X}}$ from measurements $\mathbf{Y} = \mathbf{\Phi}\mathbf{X}$ using a greedy search.

In $q$-thresholding, we select a set $\Omega$ of $k$ indices whose $q$-correlation with $\mathbf{Y}$ are among the $k$ largest. Let $\theta_k$ be the $k$th largest $q$-correlation of any $\boldsymbol{\phi}_i$ with $\mathbf{Y}$. Then we have:

$$\Omega = \{i : \|\boldsymbol{\phi}_i^T \mathbf{Y}\|_q \geq \theta_k\} \tag{15}$$

After the support $\Omega$ is determined, the non-zero coefficients of $\hat{\mathbf{X}}$ are computed via an orthogonal projection: $\hat{\mathbf{X}}_\Omega = \mathbf{\Phi}_\Omega^\dagger \mathbf{Y}$.

The $q$-simultaneous OMP (SOMP) algorithm is an iterative procedure where in each iteration, an atom is selected, and a residual is updated. The next selected atom is the one which maximizes the $q$-correlation with the current residual. The pseudocode for SOMP is summarized in Algorithm 1.

---

**Algorithm 1** Simultaneous Orthogonal Matching Pursuit (SOMP)

---

1: **initialization:** $\mathbf{R}^{(0)} = \mathbf{Y}, \mathbf{X}^{(0)} = \mathbf{0}, \Omega^0 = \emptyset$

2: **for** $n = 1; n := n + 1$ **until stopping criterion do**

3:   $i^n = \arg\max_i \|\boldsymbol{\phi}_i^T \mathbf{R}^{(n-1)}\|_q$

4:   $\Omega^n = \Omega^{n-1} \cup i^n$

5:   $\mathbf{X}_{\Omega^n,:}^{(n)} = \boldsymbol{\Phi}_{\Omega^n}^{\dagger} \mathbf{Y}$

6:   $\mathbf{R}^{(n)} = \mathbf{Y} - \boldsymbol{\Phi}\mathbf{X}^{(n)}$

7: **end for**

---

Different authors have advocated the use of different values for $q$ in step 3. However, it has been shown [11] that, independent of $q$, SOMP will recover a joint sparse representation with joint support $\Omega$ whenever the Exact Recovery Condition (ERC) [7] is met:

$$\max_{j \notin \Omega} \|\boldsymbol{\Phi}_\Omega^\dagger \boldsymbol{\phi}_j\|_1 < 1. \tag{16}$$

For the SMV problem Tropp showed that the ERC is also necessary to guarantee recovery for all vectors with support $\Omega$ [7, Theorem 3.10]. It turns out the that the necessary condition for SOMP also approaches the ERC, and furthermore, this is independent of the rank of $\mathbf{X}$. This implies that the SOMP algorithm is not able to exploit the rank information in order to improve the recovery ability in the worst-case, as we show in the theorem below.

**Theorem 3** (SOMP is not rank aware). *Let $\tau$ be given such that $1 \leq \tau \leq k$ and suppose that*

$$\max_{j \notin \Omega} \|\boldsymbol{\Phi}_\Omega^\dagger \boldsymbol{\phi}_j\|_1 > 1 \tag{17}$$

*for some support $\Omega$, $|\Omega| = k$. Then there exists an $\mathbf{X}$ with $\mathrm{supp}(\mathbf{X}) = \Omega$ and $\mathrm{rank}(\mathbf{X}) = \tau$ that SOMP cannot recover.*

*Proof:* Since the ERC does not hold, we know that for the SMV problem, there exists a vector $\mathbf{x}$ for which OMP will fail. Let $\mathbf{x}$ be a vector with $\mathrm{supp}(\mathbf{x}) = \Omega$ such that OMP incorrectly selects atom $j^\star \notin \Omega$ at the first step with

$$|\boldsymbol{\phi}_{j^\star}^T \boldsymbol{\Phi} x| > \max_{i \in \Omega} |\boldsymbol{\phi}_i^T \boldsymbol{\Phi} x| + \epsilon \tag{18}$$

for some $\epsilon > 0$. Let $\mathbf{X} := [\mathbf{x}, \mathbf{x}, \dots, \mathbf{x}]$ be the rank 1 matrix associated with $\mathbf{x}$ that cannot be recovered from $\mathbf{Y} = \boldsymbol{\Phi}\mathbf{X}$ by SOMP using any $q$.

We now perturb $\mathbf{X}$, to give $\tilde{\mathbf{X}} = \mathbf{X} + \mathbf{E}$, $\mathrm{supp}(\mathbf{E}) = \Omega$, $\max_j \|\boldsymbol{\phi}_j^T \boldsymbol{\Phi}\mathbf{E}\|_q \leq l^{1/q}\epsilon/2$ such that $\tilde{\mathbf{X}}$ has rank $\tau$ and define $\tilde{\mathbf{Y}} = \boldsymbol{\Phi}\tilde{\mathbf{X}}$. We now have:

$$
\begin{aligned}
\|\boldsymbol{\phi}_{j\cdot}^T \tilde{\mathbf{Y}}\|_q &\geq \|\boldsymbol{\phi}_{j\cdot}^T \boldsymbol{\Phi}\mathbf{X}\|_q - \|\boldsymbol{\phi}_{j\cdot}^T \boldsymbol{\Phi}\mathbf{E}\|_q \\
&\geq l^{1/q}|\boldsymbol{\phi}_{j\cdot}^T \boldsymbol{\Phi}x| - l^{1/q}\epsilon/2 \\
&> l^{1/q}\max_{i\in\Omega}|\boldsymbol{\phi}_i^T \boldsymbol{\Phi}x| + l^{1/q}\epsilon/2 \\
&= \max_{i\in\Omega}\|\boldsymbol{\phi}_i^T \boldsymbol{\Phi}\mathbf{X}\|_q + l^{1/q}\epsilon/2 \\
&\geq \max_{i\in\Omega}\left\{\|\boldsymbol{\phi}_i^T \boldsymbol{\Phi}\mathbf{X}\|_q + \|\boldsymbol{\phi}_i^T \boldsymbol{\Phi}\mathbf{E}\|_q\right\} \\
&\geq \max_{i\in\Omega}\|\boldsymbol{\phi}_i^T \tilde{\mathbf{Y}}\|_q.
\end{aligned}
\tag{19}
$$

Line 1 in (19) is the reverse triangle inequality, line 2 follows from the definition of the $q$ norm and the fact that $\mathbf{X}$ is rank 1, and line 3 follows from (18). Lines $4-6$ are a result of reversing these arguments.

Equation (19) therefore shows that no correct atom will be selected at the first step in the perturbed problem and $\tilde{\mathbf{X}}$ will not be correctly recovered. ∎

We conclude from Theorem 3 that SOMP is effectively blind to the rank of $\mathbf{X}$. An identical argument can also be used to show that an MMV version of any similar Matching Pursuit type algorithm (e.g. M-MP, M-ORMP [10]) will also be rank blind, including identification of the support set by thresholding $\|\boldsymbol{\phi}_j^T \mathbf{Y}\|_q$ as proposed in [27].

### B. Mixed $\ell^1/\ell^q$ minimization

Another popular joint sparse recovery algorithm is to perform mixed norm minimization:

$$
\hat{\mathbf{X}} = \underset{\mathbf{X}}{\operatorname{argmin}} \|\mathbf{X}\|_{1,q} \text{ s.t. } \boldsymbol{\Phi}\mathbf{X} = \mathbf{Y}
\tag{20}
$$

for some $q \geq 1$ (values of $q = 1, 2$ and $\infty$ have been advocated). This is a simple extension of the $\ell^1$ minimization used to solve SMV problems. In SMV the necessary and sufficient condition for the recovery of vectors $\mathbf{x}$ with support $\Omega$ is given by the *Null Space Property* (see [31], [37], [38]):

$$
\|z_\Omega\|_1 < \|z_{\Omega^c}\|_1, \quad \forall z \in \mathcal{N}(\boldsymbol{\Phi}).
\tag{21}
$$

Here $\Omega^c$ is the complement of the set $\Omega$.

As with SOMP we can leverage the SMV conditions for recovery to show that mixed norm methods are not rank aware:

**Theorem 4** ($\ell^1/\ell^q$ minimization is not rank aware). *Let $\tau$ be given such that $1 \leq \tau \leq k$ and suppose that there exists a $z \in \mathcal{N}(\boldsymbol{\Phi})$ such that*

$$
\|z_\Omega\|_1 > \|z_{\Omega^c}\|_1
\tag{22}
$$

*for some support $\Omega$, $|\Omega| = k$. Then there exists an $\mathbf{X}$ with $\operatorname{supp}(\mathbf{X}) = \Omega$, $\operatorname{rank}(\mathbf{X}) = \tau$ that (20) cannot recover.*

*Proof:* The proof follows along the same lines as Theorem 3. Let $\mathbf{X} = [\mathbf{x}, \mathbf{x}, \ldots, \mathbf{x}]$ be the rank 1 matrix such that $\ell^1$ minimization fails to recover $x$. Denote the $\ell^1$ minimum solution by $\hat{\mathbf{x}}$ and define $\hat{\mathbf{X}} = [\hat{\mathbf{x}}, \hat{\mathbf{x}}, \ldots, \hat{\mathbf{x}}]$.

 

Suppose that

$$||\mathbf{X}||_{1,q} > ||\hat{\mathbf{X}}||_{1,q} + \epsilon \tag{23}$$

for some $\epsilon > 0$. We now perturb $\mathbf{X}$ by a matrix $\mathbf{E}$, $\tilde{\mathbf{X}} = \mathbf{X} + \mathbf{E}$, such that $\mathrm{supp}(\mathbf{E}) = \Omega$, $\mathrm{rank}(\tilde{\mathbf{X}}) = \tau$ and $||\mathbf{E}||_{1,q} \leq \epsilon/2$. Therefore, the solution $\hat{\tilde{\mathbf{X}}}$ of (20) for the perturbed problem must satisfy:

$$||\hat{\tilde{\mathbf{X}}}||_{1,q} \leq ||\hat{\mathbf{X}}||_{1,q} + ||\mathbf{E}||_{1,q} \leq ||\hat{\mathbf{X}}||_{1,q} + \epsilon/2 < ||\mathbf{X}||_{1,q} - \epsilon/2. \tag{24}$$

On the other hand, by the triangle inequality,

$$||\tilde{\mathbf{X}}||_{1,q} \geq ||\mathbf{X}||_{1,q} - ||\mathbf{E}||_{1,q} \geq ||\mathbf{X}||_{1,q} - \epsilon/2. \tag{25}$$

Therefore, $||\tilde{\mathbf{X}}||_{1,q} > ||\hat{\tilde{\mathbf{X}}}||_{1,q}$ so (20) fails to recover the correct solution. ∎

The argument in the proof of the theorem can be applied to any joint sparse recovery that uses $||\cdot||_{p,q}$ for $p < 1$, using the appropriate Null Space Property associated with the $p$-quasi-norm [37]. Thus the result also applies to the M-FOCUSS family of algorithms [10] that define a specific means of (locally) minimizing the $\{p, q\}$-quasi-norms.

We conclude this section by noting that both greedy methods, and mixed norm approaches, are not rank aware: in the worst-case, their recovery conditions are the same as the SMV problem so that the rank cannot be exploited to improve worst-case performance. In the next section we will develop new classes of algorithms that are rank aware: in the full rank case they guarantee recovery from $k + 1$ measurements, and in the non full rank case, their performance degrades gracefully with the rank.

## VI. RANK AWARE MMV ALGORITHMS

In the last section we noted that two classes of popular techniques for joint sparse recovery were effectively rank blind. We now examine methods that can be shown to be rank aware. In particular, we consider two algorithms, both of which can be categorized as "greedy", and are discrete versions of techniques already in existence in the array processing community. The first is based on the MUSIC algorithm, and results in rank aware thresholding. The second is an extension of OMP, that is rank aware. As in the SMV problem, the thresholding approach is simpler to implement, however the OMP based approach leads to improved recovery.

### A. Rank Aware Thresholding

The simplest SMV sparse recovery algorithm selects the $i$th atom, $i \in \Omega$ if $|\boldsymbol{\phi}_i^T \mathbf{y}|$ is greater than some threshold, $\theta$ and then recovers an estimate of $\mathbf{x}$ using the pseudo-inverse, $\hat{\mathbf{x}} = \boldsymbol{\Phi}_\Omega^\dagger \mathbf{y}$. While this idea was extended to the MMV problem in [27] by considering the thresholded $q$-norm, $||\boldsymbol{\phi}_i^T \mathbf{Y}||_q$, we noted above, that as with SOMP, such an approach is not rank aware. In contrast, we have seen a rank aware version of thresholding selection in Section IV-B for the full rank case, in the form of the discrete version of MUSIC. Although it was stated there for the case $\mathrm{rank}(\mathbf{Y}) = k$ we can equally use the algorithm when $\mathrm{rank}(\mathbf{Y}) < k$, albeit with reduced performance. The idea is surprisingly simple: instead of considering $||\boldsymbol{\phi}_i^T \mathbf{Y}||_q$, we replace $\mathbf{Y}$ by an orthonormal basis for $\mathcal{R}(\mathbf{Y})$, resulting in $||\boldsymbol{\phi}_i^T \mathbf{U}||_2$ where $\mathbf{U}$ is chosen to have orthonormal columns such that $\mathcal{R}(\mathbf{U}) = \mathcal{R}(\mathbf{Y})$. The pseudocode

                                                                 

---

**Algorithm 2** Rank Aware Thresholding

---

1: Calculate $\mathbf{U} = \mathrm{orth}(\mathbf{Y})$ an orthonormal basis for $\mathcal{R}(\mathbf{Y})$;

2: Calculate $\Omega = \{i : \|\boldsymbol{\phi}_i^T \mathbf{U}\|_2 \geq \theta_k\}$ where the threshold, $\theta_k$, is set to select the $k$ largest values;

3: $\hat{\mathbf{X}} = \boldsymbol{\Phi}_\Omega^\dagger \mathbf{Y}$.

---

is summarized in Algorithm 2. When $k$ is unknown a fixed threshold, $\theta$, can be applied to $\|\boldsymbol{\phi}_i^T \mathbf{U}\|_2$ to estimate the support.

Note that, when $\mathrm{rank}(\mathbf{Y}) = 1$, rank aware thresholding reduces to standard thresholding. However, in the higher rank scenarios, unlike the standard joint thresholding techniques [27], the $\ell^2$-norm is the natural choice as this is the only norm that is invariant to the (arbitrary) choice of orthonormal basis, $\mathbf{U}$ used to represent $\mathcal{R}(\mathbf{Y})$. In the full rank case, Algorithm 2 is identical to MUSIC.

The key full rank property of this algorithm that we discussed in section IV-B was first proved in [29] and is well known in array signal processing [39]. We summarize this result in our notation.

**Theorem 5** (Feng [29]). *Let* $\mathbf{Y} = \boldsymbol{\Phi} \mathbf{X}$ *with* $|\mathrm{supp}(\mathbf{X})| = k$, $\mathrm{rank}(\mathbf{X}) = k$ *and* $k < \mathrm{spark}(\boldsymbol{\Phi}) - 1$. *Then* Rank Aware Thresholding *is guaranteed to recover* $\mathbf{X}$ *(i.e.* $\hat{\mathbf{X}} = \mathbf{X}$*).*

A direct consequence of the theorem is that $m = k + 1$ measurements are sufficient to recover $\mathbf{X}$, as long as $\boldsymbol{\Phi}$ has full spark, namely, that all sets of $k + 1$ columns are linearly independent.

*1) Practical subspace estimation:* As noted in [29], [34], due to the stability of invariant subspaces [40] the MUSIC based recovery is also robust to noise and other perturbations, though to the best of our knowledge the degree of robustness in the MMV context has not yet been fully quantified.

This, of course requires Algorithm 2 to be modified to consider the practical issue of estimating the rank of the sparse component and its signal subspace in the presence of measurement and/or numerical error. In this case we have $\mathbf{Y} = \boldsymbol{\Phi} \mathbf{X} + \mathbf{N}$ where $\mathbf{N}$ is some measurement error. The estimation of the signal subspace can be performed through a truncation of the eigenvalue decomposition or SVD of $\mathbf{Y} \mathbf{Y}^T$. It is usual to truncate the dimension of the subspace based upon the eigenvalues to distinguish between the signal and the noise subspaces. In classical MUSIC this truncation is linked to the estimation of the number of targets (equivalent to the sparsity in our model), however, when Algorithm 2 is used in the reduced rank setting it is necessary to divorce the estimated rank of the signal subspace from the sparsity level $k$.

If $\mathbf{N}$ is considered as a deterministic error term such truncation can be done based upon a known bound for $\mathbf{N} \mathbf{N}^T$. An alternative approach is to treat $\mathbf{N}$ as stochastic. This is the approach usually favoured in array processing where the problem has received considerable attention; see for example [33]. Numerous techniques have been proposed including sequential hypothesis tests, information based criteria and eigenvector detection tests. As the focus of this paper is not on the noisy scenario we do not consider these in more detail here and instead point the interested reader to [33] and the references therein.

## B. Rank Aware Pursuits

Thresholding techniques can generally be refined through the use of pursuit based methods. These involve iteratively selecting a single atom at a time, calculating an approximate solution for $\mathbf{X}$ and the residual $\mathbf{R}$, and then selecting a new atom. This is repeated until completion. The exact form of the pursuit algorithm depends on the selection strategy and the refinement step.

Here we introduce a selection strategy based upon the Rank Aware thresholding approach above.

**Rank Aware Selection** Given a residual $\mathbf{R}^{(n-1)}$ at the $(n-1)$th iteration we define the following selection strategy at the $n$th iteration:

$$\Omega^{(n)} = \Omega^{(n-1)} \cup \operatorname*{argmax}_i ||\boldsymbol{\phi}_i^T \mathbf{U}^{(n-1)}||_2, \tag{26}$$

where $\mathbf{U}^{(n-1)} = \operatorname{orth}(\mathbf{R}^{(n-1)})$.

As with the rank aware thresholding, we restrict our attention to the 2-norm since this is invariant to the orthonormal basis, $\mathbf{U}$.

The main difference between rank aware selection and standard OMP type algorithms, is that the criterion is based on the inner products with $\mathbf{U}^{(n)}$, rather than $\mathbf{R}^{(n)}$. This is the same idea that we used in rank aware thresholding: instead of comparing with the given measurements $\mathbf{Y}$ or the resulting residuals $\mathbf{R}^{(n)}$, we compare with an orthonormal basis for the span. This simple difference, which is easy to implement in practice, is enough to allow for rank awareness. Furthermore, as we will see in the simulations section, it results in enhanced performance.

Below, we introduce two algorithms that use the rank aware thresholding principle. The first is a natural generalization of SOMP, where we simply use the rank aware selection step. Although we will see that this substitution improves the performance considerably, it does not result in full rank awareness. In the full rank case, perfect recovery from $k+1$ measurements under the spark condition is not guaranteed (or typically observed as we will see in Section VII). As we show, this is a result of rank degeneration of the residual: the rank of the residual generally decreases in each iteration. To rectify this, we propose a modified selection step which forces the sparsity of the residual to decrease along with its rank. This ensures that the residual has full rank at each iteration, and results in a fully rank aware OMP-type algorithm. In simulations, we will see that this approach tends to have the best performance for all choices of the rank of $\mathbf{X}$.

Similar to other MP techniques, successful selection of an atom $\boldsymbol{\phi}_i$ from the correct support set $i \in \Omega$ requires the following condition:

$$\frac{\max_{j \notin \Omega} ||\boldsymbol{\phi}_j^T \mathbf{U}||_2}{\max_{i \in \Omega} ||\boldsymbol{\phi}_i^T \mathbf{U}||_2} < 1. \tag{27}$$

We begin by noting that for successful selection, i.e. for (27) to hold, the ERC is a sufficient but not necessary condition. Specifically we have the following proposition.

**Proposition 1.** *Let* $\mathbf{Y} = \boldsymbol{\Phi}\mathbf{X}$, $\operatorname{supp}(\mathbf{X}) = \Omega$, $|\Omega| = k$ *and* $\operatorname{rank}(\mathbf{Y}) = \tau$. *Then:*

1) $||\boldsymbol{\Phi}_\Omega^\dagger \boldsymbol{\phi}_{j^*}||_1 < 1$ *is a sufficient condition for Rank Aware selection to correctly choose an atom* $i \in \Omega$;

 

2) *If $\tau > k + 1 - |\operatorname{supp}(\mathbf{\Phi}_\Omega^\dagger \boldsymbol{\phi}_{j^\star})|$ for all $\boldsymbol{\phi}_{j^\star} := \operatorname{argmax}_{\boldsymbol{\phi}_j} \|\mathbf{\Phi}_\Omega^\dagger \boldsymbol{\phi}_j\|_1, j \notin \Omega$ then*

$$\max_{j \notin \Omega} \|\mathbf{\Phi}_\Omega^\dagger \boldsymbol{\phi}_j\|_1 \leq 1 \tag{28}$$

*is not a necessary condition for correct atom selection.*

3) *When $\tau = k$ then $k < \operatorname{spark}(\mathbf{\Phi}) - 1$ is necessary and sufficient for rank aware selection to correctly select an atom with index $i \in \Omega$.*

*Proof:* The proof of part 1 that ERC is sufficient is identical to that for SOMP, given in [11] and is based upon standard norm inequalities.

Let $\boldsymbol{\phi}_{j^\star} := \operatorname{argmax}_{\boldsymbol{\phi}_j} \|\mathbf{\Phi}_\Omega^\dagger \boldsymbol{\phi}_j\|_1, j \notin \Omega$. We can bound the norm $\|\boldsymbol{\phi}_{j^\star}^T \mathbf{U}\|_2$ for any $\mathbf{U}$ as follows:

$$\begin{aligned}
\|\boldsymbol{\phi}_{j^\star}^T \mathbf{U}\|_2 &= \max_{\mathbf{x} \neq 0} \frac{|(\mathbf{\Phi}_\Omega^\dagger \boldsymbol{\phi}_{j^\star})^T \mathbf{\Phi}_\Omega^T \mathbf{U} \mathbf{x}|}{\|\mathbf{x}\|_2} \\
&= \max_{x \neq 0} \frac{|(\mathbf{\Phi}_\Omega^\dagger \boldsymbol{\phi}_{j^\star})^T \mathbf{\Phi}_\Omega^T \mathbf{U} \mathbf{x}|}{\|\mathbf{\Phi}_\Omega^T \mathbf{U} \mathbf{x}\|_\infty} \frac{\|\mathbf{\Phi}_\Omega^T \mathbf{U} \mathbf{x}\|_\infty}{\|\mathbf{x}\|_2} \\
&\leq \|\mathbf{\Phi}_\Omega^\dagger \boldsymbol{\phi}_{j^\star}\|_1 \max_{\mathbf{x} \neq 0} \frac{\|\mathbf{\Phi}_\Omega^T \mathbf{U} \mathbf{x}\|_\infty}{\|\mathbf{x}\|_2} \\
&= \|\mathbf{\Phi}_\Omega^\dagger \boldsymbol{\phi}_{j^\star}\|_1 \max_{i \in \Omega} \|\boldsymbol{\phi}_i^T \mathbf{U}\|_2
\end{aligned} \tag{29}$$

Therefore, ERC is sufficient for recovery. However, unlike SOMP, the norm equalities cannot necessarily be approached since $\mathbf{U}$ is constrained to be column orthonormal.

For the equality in (29) we require $\mathbf{x}$ to simultaneously maximize two quantities. To maximize $\|\mathbf{\Phi}_\Omega^T \mathbf{U} \mathbf{x}\|_\infty / \|\mathbf{x}\|_2$ $\mathbf{x}$ must satisfy:

$$\mathbf{x} \propto \mathbf{U}^T \boldsymbol{\phi}_{i^\star}, \tag{30}$$

for some $\boldsymbol{\phi}_{i^\star} = \operatorname{argmax}_{\boldsymbol{\phi}_i} \|\boldsymbol{\phi}_i^T \mathbf{U}\|_2, i \in \Omega$. Secondly to maximize $|(\mathbf{\Phi}_\Omega^\dagger \boldsymbol{\phi}_{j^\star})^T \mathbf{\Phi}_\Omega^T \mathbf{U} \mathbf{x}| / \|\mathbf{\Phi}_\Omega^T \mathbf{U} \mathbf{x}\|_\infty$ we require:

$$\mathbf{\Phi}_\Omega^T \mathbf{U} \mathbf{x} \propto \operatorname{sgn}(\mathbf{\Phi}_\Omega^\dagger \boldsymbol{\phi}_{j^\star}). \tag{31}$$

Let us define the vector $\mathbf{v} \propto [\mathbf{\Phi}_\Omega^T]^\dagger \operatorname{sgn}(\mathbf{\Phi}_\Omega^\dagger \boldsymbol{\phi}_{j^\star})$, $\|\mathbf{v}\|_2 = 1$. Combining the two constraints gives:

$$\mathbf{U} \mathbf{U}^T \boldsymbol{\phi}_{i^\star} \propto \mathbf{v}. \tag{32}$$

This implies that $\mathbf{v}$ lies within the range of $\mathbf{U}$ and is the closest vector within $\mathbf{U}$ to $\boldsymbol{\phi}_{i^\star}$. However close examination of $\mathbf{v}$ shows that it is equally close to any $\boldsymbol{\phi}_i$ for which $i \in \operatorname{supp}(\mathbf{\Phi}_\Omega^\dagger \boldsymbol{\phi}_{j^\star}) \subset \Omega$. Hence all atoms $\boldsymbol{\phi}_i, i \in |\operatorname{supp}(\mathbf{\Phi}_\Omega^\dagger \boldsymbol{\phi}_{j^\star})|$ are equally as close to $\mathbf{U}$, and all must satisfy (32).

Now define the subspace $\mathcal{W}$ as:

$$\mathcal{W} := \{\mathbf{w} : \mathbf{w} = \sum_i \alpha_i \langle \boldsymbol{\phi}_i, \mathbf{v} \rangle \boldsymbol{\phi}_i, i \in \operatorname{supp}(\mathbf{\Phi}_\Omega^\dagger \boldsymbol{\phi}_{j^\star}), \sum_i \alpha_i = 0\}. \tag{33}$$

Note that by definition $\mathbf{U}^T \mathbf{w} = 0$. However by inspection we also have $\dim(\mathcal{W}) = |\operatorname{supp}(\mathbf{\Phi}_\Omega^\dagger \boldsymbol{\phi}_{j^\star})| - 1$. Since $\dim(\mathbf{U}) = \tau$ this is only possible if $\tau \leq k + 1 - |\operatorname{supp}(\mathbf{\Phi}_\Omega^\dagger \boldsymbol{\phi}_{j^\star})|$.

Thus when $\tau > k - 1 + |\operatorname{supp}(\boldsymbol{\Phi}_\Omega^\dagger \boldsymbol{\phi}_{j^\star})|$ equality in (29) cannot be achieved. Furthermore since the set of column orthonormal matrices $\mathbf{U}$ is compact a maximum must exist:

$$\max_{\mathbf{U}} \frac{\max_{j \notin \Omega} \|\boldsymbol{\phi}_j^T \mathbf{U}\|_2}{\max_{i \in \Omega} \|\boldsymbol{\phi}_i^T \mathbf{U}\|_2} = c < \max_{j \notin \Omega} \|\boldsymbol{\Phi}_\Omega^\dagger \boldsymbol{\phi}_j\|_1. \tag{34}$$

Therefore, (28) is not a necessary condition for correct atom selection, which proves part 2.

Finally when $\tau = k$ the necessary and sufficient conditions in part 3 for correct atom selection follow from the rank aware thresholding result, Theorem 5, and the identifiability conditions in Theorem 2. ∎

Proposition 1 shows that in the worst case scenario we lose nothing by incorporating the orthogonalization within the selection step. Furthermore part 2 shows that in general the selection is more effective than that of SOMP, although we have not quantified by how much as this seems difficult to estimate for general $\tau$. The exception is, of course, when $\tau = k$. In this case, RA-selection inherits the desirable property from the discrete MUSIC algorithm that correct detection is guaranteed.

*1) (Partially) Rank Aware OMP:* One possible approach to developing a Rank Aware Pursuit would be to substitute (26) as the selection step within SOMP (step 3). We call this algorithm RA-OMP. In Section VII we will see that the incorporation of the rank aware selection step substantially improves the average recovery performance over SOMP. However, curiously even when $\operatorname{rank}(\mathbf{Y}) = k$ RA-OMP is not guaranteed to exactly recover $\mathbf{X}$. That is, it does not achieve the performance of Rank Aware Thresholding and is not fully rank aware.

This can be explained by the fact that the rank of the residual deteriorates at each step of RA-OMP, a process we call *rank degeneration*. When selecting the first atom we have $\mathbf{R}^{(0)} = \mathbf{Y}$ and from Proposition 1, assuming that $\operatorname{rank}(\mathbf{Y}) = k$, we are guaranteed to select an atom $\boldsymbol{\phi}_i$ such that $i \in \Omega$. The updated residual is $\mathbf{R}^{(1)} = (I - \boldsymbol{\phi}_i \boldsymbol{\phi}_i^T)\mathbf{Y}$. Since $\mathcal{R}(\mathbf{Y}) = \mathcal{R}(\boldsymbol{\Phi}_\Omega)$, the rank will be reduced by one such that $\operatorname{rank}(\mathbf{R}^{(1)}) = k - 1$. With a little manipulation we can write $\mathbf{R}^{(1)}$ as follows:

$$\mathbf{R}^{(1)} = \sum_{j \in \Omega \setminus i} \boldsymbol{\phi}_j \mathbf{X}_{j,:} + \boldsymbol{\phi}_i \big( \sum_{j \in \Omega \setminus i} (\boldsymbol{\phi}_j^T \boldsymbol{\phi}_i) \mathbf{X}_{j,:} \big). \tag{35}$$

In general, however, the second term in (35) will not be zero and so $\mathbf{R}^{(1)}$ will still be $k$-joint sparse. This means that at the second iteration the selection step will not have guaranteed recovery since $\operatorname{rank}(\mathbf{R}^{(1)}) = k - 1$. Furthermore, if the first $\tau$ iterations have correctly selected atoms from $\Omega$. then $\operatorname{rank}(\mathbf{R}^{(\tau)}) = k - \tau$ while $\mathbf{R}^{(\tau)}$ will still generally be $k$ joint sparse. This implies that the performance of the SOMP selections will tend to degenerate towards SMV performance as the number of iterations grows.

*2) Rank Aware Order Recursive Matching Pursuit:* We can rectify the *rank degeneration* problem using a modified selection step or equivalently modifying the dictionary at each step. The idea is to force the sparsity of the residual to decrease along with its rank, so that Lemma 1 can still be applied. The mechanism that we will use for this has already been used in the SMV problem and goes by various names including: Orthogonal Least Squares [41] (since it selects the atom that minimizes the residual in the least squares sense at each step), and Order Recursive Matching Pursuit (ORMP) [3] (note that historically OMP and ORMP have been repeatedly confused for each other. For a potted history of the subject see [42]). While an MMV extension of ORMP was presented in

---

**Algorithm 3** Rank Aware Order Recursive Matching Pursuit (RA-ORMP)

---

1: **Initialize** $\mathbf{R}^{(0)} = \mathbf{Y}, \mathbf{X}^{(0)} = \mathbf{0}, \Omega^0 = \emptyset$ and $\tilde{\phi}_i = \phi_i$ for all $i$.

2: **for** $n = 1; n := n + 1$ **until stopping criterion do**

3:      Calculate orthonormal basis for residual: $\mathbf{U}^{(n-1)} = \text{Orth}(\mathbf{R}^{(n-1)})$

4:      $i^n = \text{argmax}_{i \notin \Omega^{(n-1)}} \|\tilde{\phi}_i^T \mathbf{U}^{(n-1)}\|_2$

5:      $\Omega^n = \Omega^{n-1} \cup i^n$

6:      $\mathbf{X}^{(n)}_{\Omega^n,:} = \mathbf{\Phi}^{\dagger}_{\Omega^n} \mathbf{Y}$

7:      Calculate orthogonal projector: $P^{\perp}_{\Omega^{(n)}} := (\mathbf{I} - \mathbf{\Phi}_{\Omega^{(n)}} \mathbf{\Phi}^{\dagger}_{\Omega^{(n)}})$

8:      $\mathbf{R}^{(n)} = P^{\perp}_{\Omega^{(n)}} \mathbf{Y}$

9:      $\mathbf{\Phi}' = P^{\perp}_{\Omega^{(n)}} \mathbf{\Phi}$

10:     Renormalize $\tilde{\phi}_i = \phi'_i / \|\phi'_i\|_2$, for $i \notin \Omega^{(n-1)}$.

11: **end for**

---

[10], it is based on similar norm extensions to those used in SOMP and therefore is similarly rank blind. Here we present a Rank Aware Order Recursive Matching Pursuit (RA-ORMP), and show that it has guaranteed recovery in the full rank case, as well as empirically exhibiting improved performance for all values of rank of $\mathbf{X}$. The pseudocode for RA-ORMP is given in Algorithm 3.

In practice, we do not calculate the projections as detailed above and instead use a Gramm-Schmidt orthogonalization procedure as in the standard implementation of ORMP [42]. Furthermore, a practical implementation of RA-ORMP would also need to incorporate a estimate of the signal subspace in step 3 of Algorithm 3 along the same lines as that proposed in section VI-A1.

Note the key difference between RA-OMP and RA-ORMP is that we not only project the observed data $\mathbf{Y}$ orthogonal to the selected atoms to calculate the residual, we also project the remaining dictionary atoms (step 9) and then, crucially renormalize them (step 10), so that all atoms are again unit norm.

We next show that, like RA-thresholding, RA-ORMP will exactly recover $\mathbf{X}$ in the full rank scenario:

**Theorem 6** (RA-ORMP, full rank case). *Let* $\mathbf{Y} = \mathbf{\Phi}\mathbf{X}$ *with* $|\text{supp}(\mathbf{X})| = k$, $\text{rank}(\mathbf{X}) = k$ *and* $k < \text{spark}(\mathbf{\Phi}) - 1$. *Then RA-ORMP is guaranteed to recover* $\mathbf{X}$ *(i.e.* $\hat{\mathbf{X}} = \mathbf{X}$).

*Proof:* From the spark condition we know that $\phi_j \notin \mathcal{R}(\mathbf{\Phi}_\Omega)$ for $j \notin \Omega$ and thus $\mathbf{X}$ is identifiable. From Proposition 1 and the rank assumption, the selection at the first step is successful. It is therefore sufficient to show that the updated residual provides another full rank identifiable problem.

Suppose that we select index $i \in \Omega$ at the first step. The new residual is then:

$$\mathbf{R}^{(1)} = P^{\perp}_{\Omega^{(n)}} \mathbf{Y}. \tag{36}$$

Therefore $\text{rank}(\mathbf{R}^{(1)}) = k - 1$. If we now expand $\mathbf{Y}$ in terms of the atoms $\boldsymbol{\phi}_j$ we get:

$$\mathbf{R}^{(1)} = \sum_{j \in \Omega \setminus i} P_{\Omega^{(n)}}^{\perp} \boldsymbol{\phi}_j \mathbf{X}_{j,:} \qquad (37)$$

$$= \sum_{j \in \Omega \setminus i} \tilde{\boldsymbol{\phi}}_j \tilde{\mathbf{X}}_{j,:} \qquad (38)$$

where $\tilde{\mathbf{X}}_{j,:}$ is the $j$th row of $\mathbf{X}$ rescaled by $\|P_{\Omega^{(n)}}^{\perp} \boldsymbol{\phi}_j\|_2^{-1}$. Thus $\mathbf{R}^{(1)}$ has a $(k-1)$ sparse representation within the modified dictionary $\tilde{\boldsymbol{\Phi}}$.

Finally we need to show that $\tilde{\mathbf{X}}$ is also identifiable. Since $\boldsymbol{\phi}_j \notin \mathcal{R}(\boldsymbol{\Phi}_\Omega)$ for $j \notin \Omega$, we know that $\|P_\Omega^{\perp} \boldsymbol{\phi}_j\|_2 > 0$. However this property is preserved under the projection of the dictionary since $P_\Omega^{\perp} P_i^{\perp} = P_\Omega^{\perp}$ for all $i \in \Omega$, so that $\tilde{\mathbf{X}}$ is also identifiable. Recursively applying the above arguments until $\mathbf{R}^{(k)} = 0$ completes the proof. ∎

We see that RA-ORMP does not suffer the same rank degeneration that RA-OMP does and provides a natural rank aware algorithm that reduces to ORMP in the SMV case while achieving guaranteed recovery in the full rank MMV case when $\text{rank}(\mathbf{Y}) = k$.

ORMP is usually championed as superior to OMP, since at each step it selects the atom that most decreases the size of the residual. Furthermore, empirical evidence suggests that ORMP generally outperforms OMP (slightly) but at the expense of additional computation. However, to our knowledge, there is no study of the necessary and sufficient recovery conditions for ORMP. It is easy to see that the ERC provides a necessary condition since in the first selection step OMP and ORMP are identical. Curiously though, it is not clear whether the ERC condition holding for $\boldsymbol{\Phi}$ implies that it will also hold for the modified dictionary $\tilde{\boldsymbol{\Phi}}$.

### C. Link with sequential MUSIC techniques

Within the array processing literature a number of sequential variants of MUSIC have previously been proposed [43], [44], [45], which relate to our derivations here. In [43] a sequential MUSIC algorithm is introduced which can be thought of as a continuous parameter version of a Rank Aware Matching Pursuit, i.e. RA-OMP, but without the orthogonalization step. In [45] an algorithm with the painful name of *Recursively Applied and Projected (RAP) MUSIC* was introduced. This can be thought of as a continuous parameter version of RA-ORMP. However, because in array processing the aim is to estimate a continuous parameter vector associated with the directions of arrival for multiple sources the type of analysis performed on these algorithms is very different to the exact recovery results presented here for the discrete sparsity model.

## VII. NUMERICAL EXPERIMENTS

In this section we explore the empirical performance of RA-thresholding, RA-OMP and RA-ORMP. We contrast these with results for the rank blind recovery algorithm: SOMP (using $p = 2$). For comparisons with mixed $\ell^1/\ell^q$ norm minimization we point the reader to the comparisons performed in [13], where empirically SOMP generally exhibited superior performance.

To test the four algorithms we consider the family of random matrices $\mathbf{\Phi}$ whose elements were drawn independently from a normal distribution: $\mathbf{\Phi}_{i,j} \sim \mathcal{N}(0,1)$. The columns of $\mathbf{\Phi}$ were then normalized so that $\|\boldsymbol{\phi}_i\|_2 = 1$. The dimensions of $\mathbf{\Phi}$ were fixed to $n = 256$ and $m = 32$, while the number of measurement vectors, $l$, was varied between $1$ and $32$. Finally the non-zero entries of $\mathbf{X}$ were also draw independently from a unit variance normal distribution. Note this last condition immediately implies that, with probability one, $\mathrm{rank}(\mathbf{Y}) = l$.

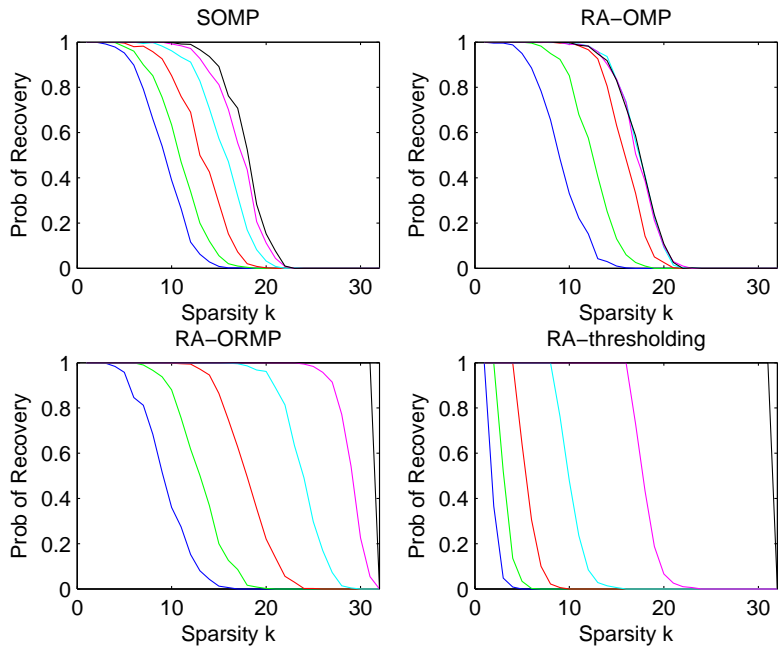

Fig. 1. The empirical probability of recovery for SOMP, RA-OMP, RA-ORMP and RA-thresholding as a function of sparsity level $k$. The curves in each plot relate to $l = 1, 2, 4, 16$ and $32$ (from left to right).

Plots of the empirical probability of recovery for the four algorithms are given in Figure 1. It is clear from these plots that while RA-OMP appears to be superior to SOMP for moderate numbers of measurement vectors ($l = 2, 4$) both SOMP and RA-OMP appear to stall at around the same sparsity level and fail to recover vectors with $k > 11$.

In contrast to this, both RA-ORMP and RA-thresholding demonstrate full recovery in the full rank case, including when $l = 32$ up to a sparsity level of $k = 31$ as predicted by the theory. However RA-thresholding does not appear to provide recovery much beyond the full rank condition. That is, recovery drops immediately once $k > l$. As with the SMV thresholding, better performance is achievable if we restrict the dynamic range of the nonzero coefficients.

The performance of RA-ORMP clearly provides the best recovery. It uniquely appears to be able to achieve OMP type performance in the SMV case and consistently provides recovery well beyond the full rank case, For example when $l = 16$ correct recovery is maintained up to $k = 23$.

In Figure 2 we show explicitly the improvement in recovery performance as a function of $l$ for the same setup as above with a fixed sparsity level of $k = m/2 = 16$. Note that for $l \leq 16$ the rank of $\mathbf{Y}$ is equal to $l$, while for $l > 16$ the rank remains constant at 16. For this level of sparsity none of the algorithms achieve significant recovery rates in the single measurement vector case. However, as the number of measurements and the rank of $\mathbf{Y}$ is increased all algorithms get improved recovery. The figure highlights that the rank aware algorithms are clearly able to exploit the rank information and the recovery rate grows to 100% when the data matrix $\mathbf{Y}$ achieves maximal rank. What is particularly striking though is how quickly the rank information improves the recovery rate in the RA-ORMP algorithm even when the rank of $\mathbf{Y}$ is significantly below the maximal rank.

In contrast, the recovery rates for SOMP and RA-OMP (recall this is not fully rank aware) do not reach 100% ever. Interestingly these plots suggest that the rank information is still playing some role in the recovery performance of SOMP as the dominant increase in its performance occurs during the range of increasing rank. When $l > 16$ and the rank remains fixed the performance appears to plateau at around 80% recovery. Additional simulations (not shown) indicate that further increasing the number of measurement vectors does not enable SOMP to pass this bound. Curiously, although the performance of RA-OMP increases much more rapidly than that of SOMP at first, it appears to stall at the same recovery rate as that for SOMP, suggesting that the rank degeneration identified in Section VI-B introduces a bottleneck to the recovery performance for RA-OMP.

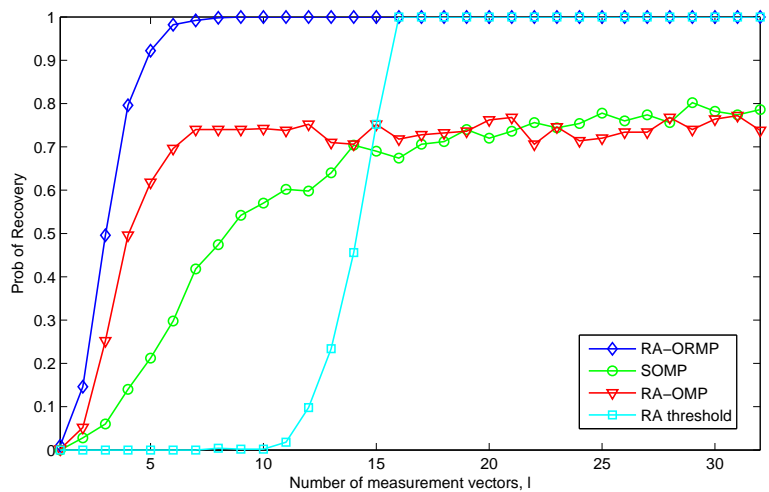

Fig. 2. The empirical probability of recovery for SOMP, RA-OMP, RA-ORMP and RA-thresholding as a function of number of measurement vectors for a fixed sparsity level, $k = m/2 = 16$.

## VIII. DISCUSSION

In this paper we considered the role of the rank of the coefficient matrix in the sparse recovery problem with multiple measurement vectors (MMV). We began by reviewing known results for the sufficient conditions for identifiability of the MMV problem and showed that these conditions are also necessary. We then demonstrated that algorithmically the rank information can be used to reduce the complexity of the general combinatorial search procedure.

Next we turned our attention to practical algorithms. The MMV problem has already received considerable study. However we have shown that the most popular classes of algorithm, $q$-SOMP and mixed $\ell^1/\ell^q$ minimization, are both rank blind. Similar arguments apply to related algorithms (e.g. mixing $\ell^p/\ell^q$ minimization) but we have omitted the details. Indeed, to our knowledge, none of the existing popular algorithms being used for joint sparse recovery are rank aware. This seemed to be a serious shortcoming of such techniques.

We therefore developed some rank aware greedy algorithms and derived certain worst case recovery conditions for these algorithms. One might (rightly) criticise such analysis as overly pessimistic however we stress that, in the full rank case, the worst case performance of RA-thresholding and RA-ORMP still significantly outperforms the average case performance for the most popular (rank blind) algorithms [13]. Such are the benefits of exploiting the rank information.

In this paper we have focused on greedy rank aware algorithms. This is because we were unable to formulate a convex rank aware algorithm. Indeed an interesting open question is: *does a convex rank aware recovery algorithm exist that can interpolate between $\ell_1$ minimization when $l = 1$ and guaranteed recovery when* $\mathrm{rank}(\mathbf{Y}) = k$? Similarly we could ask whether other popular sparse recovery algorithms such as CoSaMP [8] or Iterative Hard Thresholding [46], [47] can be adapted to create new rank aware recovery algorithms for the MMV problem.

In future work we plan to analyze the recovery properties of the proposed algorithms in more detail for the region $1 < \mathrm{rank}(\mathbf{X}) < k$. It is also important to quantify the typical performance of the algorithms rather than the worst case scenario and to account for the effects of noise. We expect an analysis similar to that in [27], [13] would be appropriate here.